\documentclass[11pt,thmsa]{article}
\usepackage{graphicx}
\usepackage[reqno,intlimits]{amsmath}
\usepackage{amsfonts}
\usepackage{amssymb}

\begin{document}

\begin{center}
\bigskip\bigskip

\bigskip

{\LARGE Weyl-Underhill-Emmrich quantization}

{\LARGE and the Stratonovich-Weyl quantizer}
\end{center}

\medskip

\bigskip

\begin{center}
Jerzy F. Pleba\'{n}ski$^{\ast}\footnote{E-mail: pleban@fis.cinvestav.mx}$,
Maciej Przanowski$^{\ast,\ast\ast}\footnote{E-mail: przan@fis.cinvestav.mx}$

Francisco J. Turrubiates$^{\ast}\footnote{E-mail: fturrub@fis.cinvestav.mx}$

\bigskip

$^{\ast}$Department of Physics

Centro de Investigaci\'{o}n y de Estudios Avanzados del IPN

Apartado Postal 14-740, M\'{e}xico, D.F., 07000, M\'{e}xico.

\smallskip

$^{\ast\ast}$Institute of Physics

Technical University of \L\'{o}d\'{z},

W\'{o}lcza\'{n}ska 219, 93-005, \L\'{o}d\'{z}, Poland.

\bigskip

\bigskip

\textbf{Abstract}
\end{center}

Weyl-Underhill-Emmrich (WUE) quantization and its generalization are
considered. It is shown that an axiomatic definition of the Stratonovich-Weyl
(SW)\ quantizer leads to severe difficulties. Quantization on the cylinder
within the WUE formalism is discussed.\bigskip

\medskip

\noindent PACS numbers: 03.65.Ca

\noindent Keywords: Quantization on Riemannian manifolds, Deformation quantization.

\begin{center}
\newpage
\end{center}

\section{Introduction}

Deformation quantization introduced in 1978 by Bayen, Flato, Fronsdal,
Lichnerowicz and Sternheimer \cite{Bayen} seems now to be one of the most
interesting part of the mathematical physics, especially after works of
Fedosov \cite{Fedosov1, Fedosov2} and Kontsevich \cite{Kont} have been
published. From the physical point of view the important question is if the
mathematical formalism of deformation quantization describes the physical
reality. One way to deal with this problem is looking for the ''natural''
generalization of the Weyl-Wigner-Moyal formalism to a Riemannian
configuration space and then comparing this with the general theory of
deformation quantization. Perhaps the most natural generalization of the Weyl
quantization rule \cite{Bayen}, [5-9] was given by Underhill \cite{Under} and
Emmrich \cite{Emm}.

In section 2 of our paper we deal with Weyl-Underhill-Emmrich (WUE) approach
and some its generalization. Then we consider how this approach leads to the
definition of Stratonovich-Weyl (SW)\ quantizer. This quantizer is used by
some authors [8,12-15] as the fundamental object defining the deformation
quantization. In our paper it is argued that the axiomatic approach to the SW
quantizer seems to lead to severe difficulties (see also \cite{Pleb2}).

In section 3 some aspects of deformation quantization on the cylinder within
the WUE formalism are considered. It is shown how in this formalism one can
define the discrete SW quantizer given by Mukunda \cite{Mukunda} and then also
obtained in \cite{Pleb2, Kasp, Olmo1}.

\section{WUE quantization and its generalization}

First assume that the configuration space of a dynamical system is the
Euclidean manifold $\mathbb{R}^{n}.$ Then the phase space is $\mathbb{R}^{2n}
$ with the natural symplectic form
\begin{equation}
\omega=d\mathbf{p}_{\alpha}\wedge d\mathbf{x}^{\alpha},\qquad\alpha
=1,...,n\tag{2.1}\label{2.1}%
\end{equation}
where $\mathbf{x}^{1},...,\mathbf{x}^{n}$ are the Cartesian coordinates on
$\mathbb{R}^{n}$ and $\mathbf{p}_{1},...,\mathbf{p}_{n}$ denote the respective momenta.

According to the Weyl quantization rule [1,5-9] if $f=f(\mathbf{p}%
,\mathbf{x})$ is a function on $\mathbb{R}^{2n}$ then the corresponding
operator $\widehat{f}_{W}$ in the space of quantum states $\mathcal{H}$ is
given by
\begin{equation}
\widehat{f}_{W}:=\int\limits_{\mathbb{R}^{2n}}\frac{d\mathbf{p}d\mathbf{x}%
}{\left(  2\pi\hbar\right)  ^{n}}f(\mathbf{p},\mathbf{x})\widehat{\Omega
}(\mathbf{p},\mathbf{x})\tag{2.2}\label{2.2}%
\end{equation}
where $d\mathbf{p}d\mathbf{x}:=d\mathbf{p}_{1}...d\mathbf{p}_{n}%
d\mathbf{x}^{1}...d\mathbf{x}^{n}$ and the operator valued function
$\widehat{\Omega}=\widehat{\Omega}(\mathbf{p},\mathbf{x})$ is defined by
\begin{equation}
\widehat{\Omega}=\widehat{\Omega}(\mathbf{p},\mathbf{x}):=2^{n}\int
\limits_{\mathbb{R}^{n}}d\xi\exp\left(  -\frac{2i\mathbf{p}\xi}{\hbar}\right)
\mid\mathbf{x}-\xi\rangle\langle\mathbf{x}+\xi\mid,\quad\mathbf{p}%
\xi:=\mathbf{p}_{\alpha}\xi^{\alpha}\tag{2.3}\label{2.3}%
\end{equation}
$\widehat{\Omega}$ is called the \textit{Stratonovich -Weyl (SW) quantizer }[8,9,12-15]

One can quickly show that
\begin{equation}
\left\{  \widehat{\Omega}(\mathbf{p},\mathbf{x})\right\}  ^{\dagger}%
=\widehat{\Omega}(\mathbf{p},\mathbf{x})\tag{2.4}\label{2.4}%
\end{equation}%
\begin{equation}
Tr\left\{  \widehat{\Omega}(\mathbf{p},\mathbf{x})\right\}  =1\tag{2.5}%
\label{2.5}%
\end{equation}
and
\begin{equation}
Tr\left\{  \widehat{\Omega}(\mathbf{p},\mathbf{x})\widehat{\Omega}%
(\mathbf{p}^{\prime},\mathbf{x}^{\prime})\right\}  =(2\pi\hbar)^{n}%
\delta(\mathbf{p}-\mathbf{p}^{\prime})\delta(\mathbf{x}-\mathbf{x}^{\prime
})\tag{2.6}\label{2.6}%
\end{equation}
The last formula, (\ref{2.6}), enables us to find the function $f=f(p,x)$ from
its Weyl image $\widehat{f}_{W}.$ Indeed, (\ref{2.2}) and (\ref{2.6}) give
\begin{equation}
f=f(\mathbf{p},\mathbf{x})=Tr\left\{  \widehat{\Omega}(\mathbf{p}%
,\mathbf{x})\widehat{f}_{W}\right\} \tag{2.7}\label{2.7}%
\end{equation}
Given any kets $\mid\varphi\rangle,\mid\psi\rangle\in\mathcal{H}$ one gets
from (\ref{2.2}) and (\ref{2.3})
\begin{align}
\langle\varphi|\widehat{f}_{W}|\psi\rangle &  =\int\limits_{\mathbb{R}^{2n}%
}\frac{d\mathbf{p}d\mathbf{x}}{\left(  2\pi\hbar\right)  ^{n}}f(\mathbf{p}%
,\mathbf{x})\langle\varphi|\widehat{\Omega}(\mathbf{p},\mathbf{x})|\psi
\rangle,\nonumber\\
\langle\varphi|\widehat{\Omega}(\mathbf{p},\mathbf{x})|\psi\rangle &
=2^{n}\int\limits_{\mathbb{R}^{n}}d\xi\exp\left(  -\frac{2i\mathbf{p}\xi
}{\hbar}\right)  \overline{\varphi(\mathbf{x}-\xi)}\psi(\mathbf{x}%
+\xi)\tag{2.8}\label{2.8}%
\end{align}
where $\varphi(\mathbf{x})=\langle\mathbf{x}|\varphi\rangle$ and
$\psi(\mathbf{x})=\langle\mathbf{x}|\psi\rangle$ denote the Schr\"{o}dinger
representation of $|\varphi\rangle,$ and $|\psi\rangle,$ respectively, and the
overbar stands for the complex conjugation. Finally,
\begin{align}
\langle\varphi|\widehat{f}_{W}|\psi\rangle &  =\frac{1}{(\pi\hbar)^{n}}%
\int\limits_{\mathbb{R}^{2n}\times\mathbb{R}^{n}}d\mathbf{p}d\mathbf{x}d\xi
f(\mathbf{p},\mathbf{x})\nonumber\\
&  \exp\left(  -\frac{2i\mathbf{p}\xi}{\hbar}\right)  \overline{\varphi
(\mathbf{x}-\xi)}\psi(\mathbf{x}+\xi)\tag{2.9}\label{2.9}%
\end{align}
In particular, let $f$ be a monomial in momenta
\begin{equation}
f=X^{\alpha_{1}...\alpha_{m}}(\mathbf{x})\mathbf{p}_{\alpha_{1}}%
...\mathbf{p}_{\alpha_{m}}\tag{2.10}\label{2.10}%
\end{equation}
where $X^{\alpha_{1}...\alpha_{m}}(\mathbf{x})$ is a totally symmetric tensor
field on the configuration space $\mathbb{R}^{n}.$ Substituing (\ref{2.10})
into (2.9), integrating with respect to $d\mathbf{p}$ and then by parts with
respect to $d\xi$ we get
\begin{align*}
\langle\varphi|\widehat{f}_{W}|\psi\rangle &  =\frac{1}{(\pi\hbar)^{n}}%
\int\limits_{\mathbb{R}^{n}\times\mathbb{R}^{n}}d\mathbf{x}d\xi X^{\alpha
_{1}...\alpha_{m}}(\mathbf{x})\left(  -\frac{\hbar}{2i}\right)  ^{m}\\
&  \overline{\varphi(\mathbf{x}-\xi)}\psi(\mathbf{x}+\xi)\frac{\partial^{m}%
}{\partial\xi^{\alpha_{1}}...\partial\xi^{\alpha_{m}}}\left\{  (2\pi
)^{n}\delta%
\genfrac{(}{)}{}{}{2\xi}{\hbar}%
\right\}
\end{align*}%
\begin{equation}
=\left(  \frac{\hbar}{2i}\right)  ^{m}\int\limits_{\mathbb{R}^{n}}%
d\mathbf{x}X^{\alpha_{1}...\alpha_{m}}(\mathbf{x})\frac{\partial^{m}}%
{\partial\xi^{\alpha_{1}}...\partial\xi^{\alpha_{m}}}\left\{  \overline
{\varphi(\mathbf{x}-\xi)}\psi(\mathbf{x}+\xi)\right\}  _{\xi=0}\tag{2.11}%
\label{2.11}%
\end{equation}
Finally, the integration by parts brings (\ref{2.11}) to the form
\begin{align}
\langle\varphi|\widehat{f}_{W}|\psi\rangle &  =\int\limits_{\mathbb{R}^{n}%
}d\mathbf{x}\overline{\varphi(\mathbf{x})}\{\left(  \frac{\hbar}{i}\right)
^{m}\sum\limits_{k=0}^{m}\frac{1}{2^{k}}\binom{m}{k}\nonumber\\
&  \left(  \partial_{\alpha_{1}}...\partial_{\alpha_{k}}X^{\alpha_{1}%
...\alpha_{k}\alpha_{k+1}...\alpha_{m}}(\mathbf{x})\right)  \partial
_{\alpha_{k+1}}...\partial_{\alpha_{m}}\}\psi(\mathbf{x})\tag{2.12}%
\label{2.12}%
\end{align}
Consequently, the Weyl image of the monomial (\ref{2.10}) reads
\begin{equation}
\widehat{f}_{W}=\left(  \frac{\hbar}{i}\right)  ^{m}\sum\limits_{k=0}^{m}%
\frac{1}{2^{k}}\binom{m}{k}\left(  \partial_{\alpha_{1}}...\partial
_{\alpha_{k}}X^{\alpha_{1}...\alpha_{k}\alpha_{k+1}...\alpha_{m}}%
(\mathbf{x})\right)  \partial_{\alpha_{k+1}}...\partial_{\alpha_{m}}%
\tag{2.13}\label{2.13}%
\end{equation}
By the linear extension of (\ref{2.13}) one obtains the Weyl image for an
arbitrary polynomial in momenta. As it has been shown in \cite{Pleb1, Cohen,
Wolf} every operator ordering satisfying some natural axioms can be obtained
with the use of an operator of the form
\begin{equation}
A=A\left(  -\hbar\frac{\partial^{2}}{\partial\mathbf{p}_{\alpha}%
\partial\mathbf{x}^{\alpha}}\right)  =1+\sum_{k=1}^{\infty}A_{k}\cdot\left(
-\hbar\frac{\partial^{2}}{\partial\mathbf{p}_{\alpha}\partial\mathbf{x}%
^{\alpha}}\right)  ^{k},\quad A_{k}\in\mathbb{C}\tag{2.14}\label{2.14}%
\end{equation}
Given operator $A$ one defines
\begin{align}
\widehat{f}^{(A)}  &  :=\int\limits_{\mathbb{R}^{2n}}\frac{d\mathbf{p}%
d\mathbf{x}}{\left(  2\pi\hbar\right)  ^{n}}\left(  Af(\mathbf{p}%
,\mathbf{x})\right)  \widehat{\Omega}(\mathbf{p},\mathbf{x})\nonumber\\
&  =\int\limits_{\mathbb{R}^{2n}}\frac{d\mathbf{p}d\mathbf{x}}{\left(
2\pi\hbar\right)  ^{n}}f(\mathbf{p},\mathbf{x})\widehat{\Omega}^{(A)}%
(\mathbf{p},\mathbf{x})\tag{2.15}\label{2.15}%
\end{align}
where
\begin{equation}
\widehat{\Omega}^{(A)}(\mathbf{p},\mathbf{x}):=A\widehat{\Omega}%
(\mathbf{p},\mathbf{x})\tag{2.16}\label{2.16}%
\end{equation}
is called the \textit{generalized Stratonovich-Weyl quantizer }\cite{Pleb1}%
\textit{. }We have
\begin{equation}
Tr\left\{  \widehat{\Omega}^{(A)}(\mathbf{p},\mathbf{x})\right\}
=1\tag{2.17}\label{2.17}%
\end{equation}%
\begin{equation}
Tr\left\{  \widehat{\Omega}^{(A)}(\mathbf{p},\mathbf{x})\widehat{\Omega}%
^{(A)}(\mathbf{p}^{\prime},\mathbf{x}^{\prime})\right\}  =(2\pi\hbar)^{n}%
A^{2}\left(  -\hbar\frac{\partial^{2}}{\partial\mathbf{p}_{\alpha}%
\partial\mathbf{x}^{\alpha}}\right)  \delta(\mathbf{p}-\mathbf{p}^{\prime
})\delta(\mathbf{x}-\mathbf{x}^{\prime})\tag{2.18}\label{2.18}%
\end{equation}
Hence one gets the generalization of the formula (\ref{2.7})
\begin{equation}
f=f(\mathbf{p},\mathbf{x})=A^{-2}\left(  -\hbar\frac{\partial^{2}}%
{\partial\mathbf{p}_{\alpha}\partial\mathbf{x}^{\alpha}}\right)  Tr\left\{
\widehat{\Omega}^{(A)}(\mathbf{p},\mathbf{x})\widehat{f}^{(A)}\right\}
\tag{2.19}\label{2.19}%
\end{equation}
For the Weyl ordering we have
\begin{equation}
A=1\tag{2.20}\label{2.20}%
\end{equation}
and for the so called \textit{standard ordering}
\begin{equation}
A=\exp\left\{  \frac{i\hbar}{2}\frac{\partial^{2}}{\partial\mathbf{p}_{\alpha
}\partial\mathbf{x}^{\alpha}}\right\} \tag{2.21}\label{2.21}%
\end{equation}
In what follows we denote by $\widehat{f}_{W}$ and $\widehat{f}_{S}$ the Weyl
and the standard ordering, respectively. One can easily show that if $f$ is
the monomial (\ref{2.10}) then
\begin{equation}
\widehat{f}_{S}=%
\genfrac{(}{)}{}{}{\hbar}{i}%
^{m}X^{\alpha_{1}...\alpha_{m}}(\mathbf{x})\partial_{\alpha_{1}}%
...\partial_{\alpha_{m}}\tag{2.22}\label{2.22}%
\end{equation}
It is evident that $\widehat{f}^{(A)}$ is hermitian for every real monomial of
the form (\ref{2.10}) if and only if
\begin{equation}
\overline{A}=A\tag{2.23}\label{2.23}%
\end{equation}
Our intent is to generalize the above considerations on the case when the
configuration space is an n-dimensional Riemannian manifold $\left(
M,g\right)  ,$ where $g\in Symm(T^{\ast}M\otimes T^{\ast}M)$ is the metric on $M.$

The phase space is the cotangent bundle $T^{\ast}M$ over $M$ endowed with the
natural symplectic form
\begin{equation}
\omega=dp_{\alpha}\wedge dq^{\alpha}\quad\alpha=1,...,n\tag{2.24}\label{2.24}%
\end{equation}
where $q^{1},...,q^{n}$ are coordinates in $M$ and $p_{1},...,p_{n}%
,q^{1},...,q^{n}$ are the induced coordinates (the proper Darboux coordinates)
in $T^{\ast}M$.

Let $f=f(p,q)$ be a function on $T^{\ast}M$. The question is to find a natural
generalization of the Weyl quantization rule for $\mathbb{R}^{2n}$ to the case
of $T^{\ast}M.$

It seems that the best answer to this question has been done by Underhill
\cite{Under} and then by Emmrich \cite{Emm}. We follow them changing only the
measure used in the integration over $TM.$ (About the Underhill-Emmrich
approach see also distinguished papers by Bordemann, Neumaier and Waldmann
\cite{Bordem1, Bordem2} and Pflaum \cite{Pflaum1, Pflaum2}).

The first glance at the formulas (\ref{2.2}), (\ref{2.3}) and (2.9) shows that
the main problem lies in a definition of the term $\exp\left(  -\frac{2ip\xi
}{\hbar}\right)  $ when $M$ is no longer the Euclidean space $\mathbb{R}^{n}.$
In the Underhill-Emmrich approach it is done by the use of normal coordinates.

Let $q$ be any point of $M$ and $T_{q}(M)$ and $T_{q}^{\ast}(M)$ be the
tangent and cotangent space of $M$ at $q$, respectively. For any $\xi
=\xi^{\alpha}\left(  \frac{\partial}{\partial q^{\alpha}}\right)  _{q}\in
T_{q}(M)$ and $p=p_{\alpha}(dq^{\alpha})_{q}\in T_{q}^{\ast}(M)$ we write as
before $p\xi:=p_{\alpha}\xi^{\alpha}.$

For every $q\in M$ we choose a normal neighborhood $V_{q}^{\prime}\subset
T_{q}(M),$ an open ball $K_{q}\subset V_{q}^{\prime}$ and some smaller
neighborhood of $q,$ $V_{q}\subset K_{q}.$ Then one defines a cutoff function
$\chi=\chi(q,\xi)\in C^{\infty}\left(  TM\right)  $ such that for every $q\in
M$%
\begin{equation}
\chi(q,\xi)=\left\{
\begin{array}
[c]{c}%
1\,for\,\text{\ \ \ \ }\xi\in V_{q}\\
0\text{ }for\text{ \ \ \ \ }\xi\notin K_{q}%
\end{array}
\right. \tag{2.25}\label{2.25}%
\end{equation}
Let $\exp_{q}:V_{q}^{\prime}\longrightarrow U_{q}\subset M$ be the exponential
map of $V_{q}^{\prime}$ onto $U_{q}.$ For any functions $\varphi$ and $\psi$
on $M$ and for every point $q\in M$ we define the functions $\Phi_{q}^{-}$ and
$\Psi_{q}^{+}$ on $T_{q}(M)$ by
\begin{align}
\Phi_{q}^{-}(\xi)  &  =\left\{
\begin{array}
[c]{c}%
\chi(q,-\xi)\varphi(\exp_{q}(-\xi))\text{ \ \ \ \ }for\,\xi\in K_{q}\\
\text{ \ \ \ \ \ \ \ \ \ \ \ }0\text{\ \ \ \ \ \ \ \ \ \ \ \ \ \ \ \ \ \ \ }%
for\text{ }\xi\notin K_{q}%
\end{array}
\right. \nonumber\\
\Psi_{q}^{+}(\xi)  &  =\left\{
\begin{array}
[c]{c}%
\chi(q,\xi)\psi(\exp_{q}\xi)\text{ \ \ \ \ \ \ \ \ \ \ }for\,\xi\in K_{q}\\
\text{ \ \ \ \ \ \ \ \ }0\text{\ \ \ \ \ \ \ \ \ \ \ \ \ \ \ \ \ \ \ \ \ }%
for\text{ }\xi\notin K_{q}%
\end{array}
\right. \tag{2.26}\label{2.26}%
\end{align}
Let $f=f(p,q)$ be a function on $T^{\ast}M.$

By the analogy to (2.9) one assigns to $f$ the following operator $\widehat
{f}_{W}$%
\begin{align}
\langle\varphi|\widehat{f}_{W}|\psi\rangle &  :=\frac{1}{(\pi\hbar)^{n}}%
\int\limits_{T^{\ast}M}dpdqf(p,q)\int\limits_{T_{q}(M)}\sqrt{g(\xi)}d\xi
\exp\left(  -\frac{2ip\xi}{\hbar}\right) \nonumber\\
&  \overline{\Phi_{q}^{-}(\xi)}\Psi_{q}^{+}(\xi)\tag{2.27}\label{2.27}%
\end{align}
Then we have also
\begin{align}
\langle\varphi|\widehat{f}_{W}|\psi\rangle &  =\int\limits_{T^{\ast}M}%
\frac{dpdq}{(2\pi\hbar)^{n}}f(p,q)\langle\varphi\mid\widehat{\Omega}%
(p,q)\mid\psi\rangle\tag{2.28}\\
\langle\varphi|\widehat{\Omega}(p,q)|\psi\rangle &  =2^{n}\int\limits_{T_{q}%
(M)}\sqrt{g(\xi)}d\xi\exp\left(  -\frac{2ip\xi}{\hbar}\right)  \overline
{\Phi_{q}^{-}(\xi)}\Psi_{q}^{+}(\xi)\nonumber
\end{align}
where $g(\xi)$ stands for the determinant of the metric on$\ M$ in the normal coordinates.

Note that Underhill \cite{Under} assumes the measure to be $d\xi$ and,
consequently $\varphi$ and $\psi$ are half-densities. On the other hand
Emmrich \cite{Emm} deals with the measure $\sqrt{g(\xi)}d\xi$ and therefore
$\varphi$ and $\psi$ are scalars. We assume that the wave functions $\varphi$
and $\psi$ are scalars but the measure on $V_{q}^{\prime}\subset T_{q}(M)$ is
$\sqrt{g(\xi)}d\xi.$

The operator $\widehat{\Omega}(p,q)$ defined by (2.28) plays now the role of
the SW quantizer. The only problem is that both $\widehat{\Omega}$ and
$\widehat{f}_{W}$ depend on the cutoff function $\chi(q,\xi).$ Thus one should
find the ''optimal'' form of $\chi.$ However, as it was shown by Underhill
\cite{Under}, if the function $f$ is a polynomial with respect to mementa then
$\widehat{f}_{W}$ doesn't depend on $\chi.$ Indeed, let
\begin{equation}
f=f(p,q)=X^{\alpha_{1}...\alpha_{m}}(q)p_{\alpha_{1}}...p_{\alpha_{m}%
}\tag{2.29}\label{2.29}%
\end{equation}
Substituing (\ref{2.29}) into (2.27), integrating with respect to $dp$ and
then, by parts, with respect to $d\xi$ one gets
\begin{align}
\langle\varphi|\widehat{f}_{W}|\psi\rangle &  =\frac{1}{(\pi\hbar)^{n}}%
\int\limits_{T^{\ast}M}dpdqX^{\alpha_{1}...\alpha_{m}}(q)\int\limits_{T_{q}%
(M)}\sqrt{g(\xi)}d\xi\left(  -\frac{\hbar}{2i}\right)  ^{m}\nonumber\\
&  \left\{  \frac{\partial^{m}}{\partial\xi^{\alpha_{1}}...\partial\xi
^{\alpha_{m}}}\exp\left(  -\frac{2ip\xi}{\hbar}\right)  \right\}
\overline{\Phi_{q}^{-}(\xi)}\Psi_{q}^{+}(\xi)\nonumber\\
&  =\left(  \frac{\hbar}{2i}\right)  ^{m}\int\limits_{M}\sqrt{g(q)}%
dqX^{\alpha_{1}...\alpha_{m}}(q)\left\{  \frac{\partial^{m}}{\partial
\xi^{\alpha_{1}}...\partial\xi^{\alpha_{m}}}\widetilde{D}(q,\xi)\right\}
_{\xi=0}\nonumber\\
\widetilde{D}(q,\xi) &  :=\frac{\sqrt{g(\xi)}}{\sqrt{g(q)}}\overline{\Phi
_{q}^{-}(\xi)}\Psi_{q}^{+}(\xi)\tag{2.30}\label{2.30}%
\end{align}
However, it is an easy matter to show that (see Petrov \cite{Petrov})
\begin{align}
\left\{  \frac{\partial^{k}}{\partial\xi^{\alpha_{1}}...\partial\xi
^{\alpha_{k}}}\overline{\Phi^{-}(\xi)}\right\}  _{\xi=0} &  =(-1)^{k}%
\nabla_{(\alpha_{1}}...\nabla_{\alpha_{k})}\overline{\varphi(q)}\nonumber\\
\left\{  \frac{\partial^{k}}{\partial\xi^{\alpha_{1}}...\partial\xi
^{\alpha_{k}}}\Psi^{+}(\xi)\right\}  _{\xi=0} &  =\nabla_{(\alpha_{1}%
}...\nabla_{\alpha_{k})}\psi(q)\tag{2.31}\label{2.31}%
\end{align}
where $\nabla_{\alpha_{1}}:=\nabla_{\frac{\partial}{\partial q^{\alpha_{1}}}%
},...$ etc., and the bracket $\left(  \alpha_{1}...\alpha_{k}\right)  $ stands
for the symmetrization.

Finally, inserting (2.31) into (2.30) and integrating by parts one arrives at
the following result being a generalization of the one obtained by Bordemann
et al \cite{Bordem2}.
\[
\langle\varphi\mid\widehat{f}_{W}\mid\psi\rangle=\int\limits_{M}\sqrt
{g(q)}dq\overline{\varphi(q)}\widehat{f}_{W}\psi(q)
\]%
\begin{align*}
\widehat{f}_{W}  &  =%
\genfrac{(}{)}{}{}{\hbar}{i}%
^{m}\sum\limits_{k=0}^{m}\binom{m}{k}\sum_{j=0}^{m-k}\binom{m-k}{j}\frac
{1}{2^{k+j}}\\
&  \left(  \nabla_{\alpha_{1}}...\nabla_{\alpha_{j}}\widetilde{X}^{\alpha
_{1}...\alpha_{j}\alpha_{j+1}...\alpha_{m-k}}(q)\right)  \nabla_{\alpha_{j+1}%
}...\nabla_{\alpha_{m-k}}%
\end{align*}%
\begin{align*}
&  =\sum\limits_{k=0}^{m}%
\genfrac{(}{)}{}{}{\hbar}{2i}%
^{k}\binom{m}{k}\{%
\genfrac{(}{)}{}{}{\hbar}{i}%
^{m-k}\sum\limits_{j=0}^{m-k}\frac{1}{2j}\binom{m-k}{j}\\
&  \left(  \nabla_{\alpha_{1}}...\nabla_{\alpha_{j}}\widetilde{X}^{\alpha
_{1}...\alpha_{j}\alpha_{j+1}...\alpha_{m-k}}(q)\right)  \nabla_{\alpha_{j+1}%
}...\nabla_{\alpha_{m-k}}\}
\end{align*}%
\begin{align}
\widetilde{X}^{\alpha_{1}...\alpha_{j}\alpha_{j+1}...\alpha_{m-k}}(q)  &
:=X^{\beta_{1}...\beta_{k}\alpha_{1}...\alpha_{j}\alpha_{j+1}...\alpha_{m-k}%
}(q)\nonumber\\
&  \left\{  \frac{\partial^{k}}{\partial\xi^{\beta_{1}}...\partial\xi
^{\beta_{k}}}\frac{\sqrt{g(\xi)}}{\sqrt{g(q)}}\right\}  _{\xi=0}%
\tag{2.32}\label{2.32}%
\end{align}
The term corresponding to $k=0$ is exactly the operator given in
\cite{Bordem1, Bordem2}. (Compare also with (\ref{2.13})). Thus one concludes
that if $f=f(p,q)$ is a monomial of the form (\ref{2.29}) then $\widehat
{f}_{W}$ given by (2.32) is independent of the cutoff function $\chi$. By
linearity this is also true for any polynomial with respect to momenta.

\textit{Examples:} (compare with \cite{Under, Emm, Bordem2})

\noindent(i) Assume
\begin{equation}
f=X^{\alpha}(q)p_{\alpha}\tag{2.33}%
\end{equation}
then
\begin{equation}
\widehat{f}_{W}=\frac{\hbar}{i}\left[  X^{\alpha}(q)\nabla_{\alpha}+\frac
{1}{2}\left(  \nabla_{\alpha}X^{\alpha}(q)\right)  \right] \tag{2.34}%
\label{2.34}%
\end{equation}
(ii)
\begin{equation}
f=X^{\alpha\beta}(q)p_{\alpha}p_{\beta}\tag{2.35}\label{2.35}%
\end{equation}
Here
\[
\widehat{f}_{W}=\left(  \frac{\hbar}{i}\right)  ^{2}[X^{\alpha\beta}%
(q)\nabla_{\alpha}\nabla_{\beta}+\left(  \nabla_{\alpha}X^{\alpha\beta
}(q)\right)  \nabla_{\beta}+
\]%
\begin{equation}
\frac{1}{4}\left(  \nabla_{\alpha}\nabla_{\beta}X^{\alpha\beta}(q)\right)
+\frac{1}{12}X^{\alpha\beta}(q)R_{\alpha\beta}(q)]\tag{2.36}\label{2.36}%
\end{equation}
where $R_{\alpha\beta}(q)$ is the Ricci tensor on $M$%
\begin{equation}
R_{\alpha\beta}=R_{\alpha\gamma\beta}^{\gamma}=\partial_{\gamma}\Gamma
_{\alpha\beta}^{\gamma}-\partial_{\beta}\Gamma_{\alpha\gamma}^{\gamma}%
+\Gamma_{\gamma\delta}^{\gamma}\Gamma_{\alpha\beta}^{\delta}-\Gamma
_{\beta\delta}^{\gamma}\Gamma_{\alpha\gamma}^{\delta}\tag{2.37}\label{2.37}%
\end{equation}
(iii) Let
\begin{align}
f  &  =X^{\alpha\beta}(q)p_{\alpha}p_{\beta}+i\hbar\left(  \nabla_{\alpha
}X^{\alpha\beta}(q)\right)  p_{\beta}-\frac{\hbar^{2}}{4}\left(
\nabla_{\alpha}\nabla_{\beta}X^{\alpha\beta}(q)\right) \nonumber\\
&  +\frac{\hbar^{2}}{12}X^{\alpha\beta}(q)R_{\alpha\beta}(q)\tag{2.38}%
\label{2.38}%
\end{align}
Then
\begin{equation}
\widehat{f}_{W}=\left(  \frac{\hbar}{i}\right)  ^{2}X^{\alpha\beta}%
(q)\nabla_{\alpha}\nabla_{\beta}\tag{2.39}\label{2.39}%
\end{equation}
Now we are in a position to consider an important problem. As it has been
mentioned the operator $\widehat{\Omega}(p,q)$ given by (2.28) is the SW
quantizer within the WUE formalism. Of course $\widehat{\Omega}(p,q)$ depends
on the cutoff function $\chi(q,\xi).$ Therefore the question is if there
exists $\chi(q,\xi)$ such that the usual axioms of the SW quantizer [8,13-15]
i.e.
\begin{equation}
\left\{  \widehat{\Omega}(p,q)\right\}  ^{\dagger}=\widehat{\Omega
}(p,q)\tag{2.40}\label{2.40}%
\end{equation}%
\begin{equation}
Tr\left\{  \widehat{\Omega}(p,q)\right\}  =1\tag{2.41}\label{2.41}%
\end{equation}%
\[
\int\limits_{T^{\ast}M}\frac{dp^{\prime}dq^{\prime}}{(2\pi\hbar)^{n}%
}f(p^{\prime},q^{\prime})Tr\left\{  \widehat{\Omega}(p,q)\widehat{\Omega
}(p^{\prime},q^{\prime})\right\}  =f(p,q)
\]%
\begin{equation}
\Longleftrightarrow Tr\left\{  \widehat{\Omega}(p,q)\widehat{f}_{W}\right\}
=f(p,q)\tag{2.42}%
\end{equation}
are satisfied by the operator $\widehat{\Omega}(p,q)$ defined by (2.28) (see
(\ref{2.4}) to (\ref{2.7})).

It is evident that the condition (\ref{2.40}) holds for any $\chi(q,\xi).$ Now
to check (\ref{2.41}) we take a complete orthonormal system of functions
$\left\{  \varphi_{j}\right\}  $ on $M$%
\begin{align}
\int\limits_{M}\sqrt{g(q)}dq\overline{\varphi_{k^{\prime}}(q)}\varphi_{k}(q)
&  =\delta_{kk^{\prime}}\nonumber\\
\sum\limits_{k}\overline{\varphi_{k}(q^{\prime})}\varphi_{k}(q)  &
=\frac{\delta(q-q^{\prime})}{\sqrt{g(q)}}\tag{2.43}\label{2.43}%
\end{align}
It is an easy matter to observe that without any loss of generality one can
use the exponential functions in the tangent space $T_{q}(M)$%
\begin{align}
\varphi_{s}(\xi)  &  =\frac{1}{\left(  \sqrt{2\pi}\right)  ^{n}\sqrt[4]%
{g(\xi)}}\exp(is\xi)\tag{2.44}\label{2.44}\\
s  &  =\left(  s_{1},...,s_{n}\right)  \in\mathbb{Z}\times...\times
\mathbb{Z}\nonumber
\end{align}
Consequently we get
\[
Tr\left\{  \widehat{\Omega}(p,q)\right\}  =\sum\limits_{s\in\mathbb{Z}%
\times...\times\mathbb{Z}}\langle\varphi_{s}\mid\widehat{\Omega}%
(p,q)\mid\varphi_{s}\rangle
\]%
\begin{equation}
=2^{n}\int\limits_{T_{q}(M)}\sqrt{g(\xi)}d\xi\frac{\chi(q,-\xi)\chi(q,\xi
)}{\left(  2\pi\right)  ^{n}\sqrt[4]{g(-\xi)}\sqrt[4]{g(\xi)}}(2\pi)^{n}%
\delta(2\xi)=1\tag{2.45}\label{2.45}%
\end{equation}
(\textit{Remark}: In (2.44) and (2.45) it is assumed that $K_{q}\subset\left[
-\pi,\pi\right]  \times...\times\left[  -\pi,\pi\right]  .$ In other case we
should change the period of the exponential functions but the final result of
(2.45) holds true).

Thus (\ref{2.41}) is fulfilled for every cutoff funtion $\chi(q,\xi).$
Consider now the condition (2.42). To this end we use the example (iii).
Inserting the operator (\ref{2.39}) into (2.42), using as before the
exponential functions (2.44) and employing also some formulas from the theory
of the normal coordinate systems \cite{Petrov} one arrives at the following
result
\begin{equation}
f(p,q)-Tr\left\{  \widehat{\Omega}(p,q)\widehat{f}_{W}\right\}  =\frac
{\hbar^{2}}{3}X^{\alpha\beta}(q)R_{\alpha\beta}(q)\tag{2.46}\label{2.46}%
\end{equation}
where $f=f(p,q)$ is defined by (2.38). As (\ref{2.46}) holds true for an
arbitrary $\chi(q,\xi)$ the axiom (2.42) cannot be satisfied. One can quickly
show that the analogous result to (\ref{2.46}) holds true when the Emmrich
measure $\sqrt{g(q)}d\xi$ is considered. Thus we conclude that: \textit{in
general the axiom }(2.42)\textit{\ is not satisfied within the WUE formalism
for any choice of the cutoff function }$\chi(q,\xi).$ Therefore, from the WUE
formalism point of view the axiomatic approach to the definition of the SW
quantizer seems to be questionable. (See also \cite{Pleb2, Olmo1} and the next
section of the present paper).

Finally let us consider the problem of different operator orderings.

One can quickly find that in the Euclidean case if we perform a point
transformation
\begin{equation}
q^{\alpha}=q^{\alpha}\left(  \mathbf{x}^{\beta}\right)  ,\quad p_{\alpha
}=\frac{\partial\mathbf{x}^{\beta}\left(  q^{\gamma}\right)  }{\partial
q^{\alpha}}\mathbf{p}_{\beta}\tag{2.47}%
\end{equation}
where $\mathbf{x}^{1},...,\mathbf{x}^{n}$ are the Cartesian coordinates and
$\mathbf{p}_{1},...,\mathbf{p}_{n}$ corresponding momenta, then
\begin{equation}
-\hbar\frac{\partial^{2}}{\partial\mathbf{p}_{\alpha}\partial\mathbf{x}%
^{\alpha}}=-\hbar\left\{  \frac{\partial^{2}}{\partial p_{\alpha}\partial
q^{\alpha}}+p_{\gamma}\Gamma_{\alpha\beta}^{\gamma}(q)\frac{\partial^{2}%
}{\partial p_{\alpha}\partial p_{\beta}}+\Gamma_{\alpha\beta}^{\beta}%
(q)\frac{\partial}{\partial p_{\alpha}}\right\} \tag{2.48}\label{2.48}%
\end{equation}
where $\Gamma_{\beta\gamma}^{\alpha}$ are Christoffel's symbols with respect
to the coordinates $q^{\alpha}.$ Hence, it is natural to generalize the object
$A$ defining the operator ordering in the Euclidean case (see (\ref{2.14})) to
the following one
\begin{align}
A  &  =A(\Delta)=1+\sum_{k=1}^{\infty}A_{k}\Delta^{k},\quad A_{k}\in
\mathbb{C}\nonumber\\
\Delta &  :=-\hbar\left(  \frac{\partial^{2}}{\partial p_{\alpha}\partial
q^{\alpha}}+p_{\gamma}\Gamma_{\alpha\beta}^{\gamma}\frac{\partial^{2}%
}{\partial p_{\alpha}\partial p_{\beta}}+\Gamma_{\alpha\beta}^{\beta}%
\frac{\partial}{\partial p_{\alpha}}\right) \tag{2.49}\label{2.49}%
\end{align}
when the configuration space is an $n$-dimensional Riemannian manifold
$(M,ds^{2}).$ (Operator $\Delta$ was also found by Bordemann et al
\cite{Bordem1, Bordem2}).

Consequently, we have now
\begin{align}
\langle\varphi|\widehat{f}^{(A)}|\psi\rangle &  =\int\limits_{T^{\ast}M}%
\frac{dpdq}{(2\pi\hbar)^{n}}\left(  Af(p,q)\right)  \langle\varphi\mid
\widehat{\Omega}(p,q)\mid\psi\rangle\nonumber\\
&  =\int\limits_{T^{\ast}M}\frac{dpdq}{(2\pi\hbar)^{n}}f(p,q)\langle
\varphi\mid\widehat{\Omega}^{(A)}(p,q)\mid\psi\rangle\tag{2.50}\label{2.50}%
\end{align}
where the generalized SW quantizer $\widehat{\Omega}^{(A)}(p,q)$ is defined
by
\begin{equation}
\widehat{\Omega}^{(A)}(p,q):=A\widehat{\Omega}(p,q)\tag{2.51}\label{2.51}%
\end{equation}
In particular for the \textit{generalized standard ordering} one put
\cite{Bordem1}
\begin{equation}
A=\exp\left\{  \frac{i\hbar}{2}\Delta\right\} \tag{2.52}\label{2.52}%
\end{equation}
and for the monomial (2.29) we get
\begin{equation}
\widehat{f}_{S}:=\widehat{f}^{(A)}=%
\genfrac{(}{)}{}{}{\hbar}{i}%
^{m}\sum\limits_{k=0}^{m}\frac{1}{2^{k}}\binom{m}{k}\widetilde{X}^{\alpha
_{1}...\alpha_{m-k}}(q)\nabla_{\alpha_{1}}...\nabla_{\alpha_{m-k}}%
\tag{2.53}\label{2.53}%
\end{equation}
The term with $k=0$ corresponds exactly to the operator in standard ordering
in the case of the Emmrich measure \cite{Bordem1}.

\section{Quantization on the cylinder}

Consider a simple dynamical system consisting of one particle on the circle
$S^{1}.$ The phase space of this system is the cylinder $\mathbb{R\times}%
S^{1}.$ The deformation quantization for this case might seem to be a simple
modification of the Euclidean case. However, it is not because of the
non-trivial topology of $S^{1}.$ In particular one arrives at the conclusion
that if the deformation quantization on the cylinder $\mathbb{R\times}S^{1}$
is to give ''physical'' results then the classical phase space should be
quantized to be $\hbar\mathbb{Z}\times S^{1}$ [16-19]. Here we consider some
aspects of the deformation quantization on the cylinder using the
Weyl-Underhill-Emmrich quantiztion rule. In the present case the configuration
space $M=S^{1},$ then $T^{\ast}M=\mathbb{R\times}S^{1}$ and $T_{q}%
M=\mathbb{R}$. For the coordinate $q$ we use the angle $\theta$, $-\pi
\leq\theta<\pi$. The complete orthonormal system of $L^{2}(S^{1})$ is given
by
\begin{equation}
\varphi_{k}=\frac{1}{\sqrt{2\pi}}\exp(ik\theta),\quad k\in\mathbb{Z}%
\tag{3.1}\label{3.1}%
\end{equation}
For simplicity we assume that the cutoff function $\chi(\theta,\xi)$ is
symmetric with respect to $\xi$%
\begin{equation}
\chi(\theta,\xi)=\chi(\theta,-\xi)\quad\forall\,\theta\in\lbrack-\pi
,\pi\lbrack\tag{3.2}\label{3.2}%
\end{equation}
The SW quantizer $\widehat{\Omega}(p,\theta)$ defined by (2.28) reads now
\begin{align}
\langle\varphi_{k}|\widehat{\Omega}(p,\theta)|\varphi_{k^{\prime}}\rangle &
=\frac{1}{\pi}\exp\left\{  i(k^{\prime}-k)\theta\right\} \nonumber\\
&  \int\limits_{-\infty}^{\infty}d\xi\chi^{2}(\theta,\xi)\exp\left\{  i\left(
k+k^{\prime}-\frac{2p}{\hbar}\right)  \xi\right\} \tag{3.3}%
\end{align}
One can quickly check that according to the general formula (\ref{2.45})
\begin{equation}
Tr\left\{  \widehat{\Omega}(p,\theta)\right\}  =\sum\limits_{k\in\mathbb{Z}%
}\langle\varphi_{k}\mid\widehat{\Omega}(p,\theta)\mid\varphi_{k}%
\rangle=1\tag{3.4}\label{3.4}%
\end{equation}
for arbitrary $\chi.$ Let $f=f(p,\theta)$ be a monomial
\begin{equation}
f(p,\theta)=X(\theta)p^{m}\tag{3.5}\label{3.5}%
\end{equation}
then
\[
\int\limits_{\mathbb{R\times}S^{1}}\frac{dp^{\prime}d\theta^{\prime}}%
{2\pi\hbar}f(p^{\prime},\theta^{\prime})Tr\left\{  \widehat{\Omega}%
(p,\theta)\widehat{\Omega}(p^{\prime},\theta^{\prime})\right\}  =\sum
\limits_{k\in\mathbb{Z}}\langle\varphi_{k}\mid\widehat{\Omega}(p,\theta
)\widehat{f}_{W}\mid\varphi_{k}\rangle
\]%
\begin{equation}
=\int\limits_{-\infty}^{\infty}d\xi\chi^{2}(\theta,\xi)\exp\left(
-\frac{2ip\xi}{\hbar}\right)
\genfrac{(}{)}{}{}{\hbar}{2i}%
^{m}\frac{\partial^{m}}{\partial\xi^{m}}\left(  X(\theta+\xi)\delta
(\xi)\right)  =X(\theta)p^{m}\tag{3.6}\label{3.6}%
\end{equation}
By the linearity of the integral (2.42) with respect to $f$ one concludes that
\textit{the} \textit{axiom }(2.42) \textit{is now satisfied for a function}
$f=f(p,\theta)$ \textit{being an arbitrary polynomial in the momentum} $p.$

If we want the axiom (2.42) to hold for any function on the cylinder then
$Tr\left\{  \widehat{\Omega}(p,\theta)\widehat{\Omega}(p^{\prime}%
,\theta^{\prime})\right\}  $ should be equal to $2\pi\hbar\delta(\theta
-\theta^{\prime})\delta(p-p^{\prime}).$ Performing simple manipulations,
remembering also that $\chi(\theta,\xi)=0$ for $\xi\neq]-\pi,\pi\lbrack$
(i.e.$K_{\theta}\subset$ $\ ]-\pi,\pi\lbrack$) one finds
\[
Tr\left\{  \widehat{\Omega}(p,\theta)\widehat{\Omega}(p^{\prime}%
,\theta^{\prime})\right\}  =2\delta(\theta-\theta^{\prime})\int
\limits_{-\infty}^{\infty}d\xi\chi^{4}(\theta,\xi)\exp\left\{  \frac{2i}%
{\hbar}(p^{\prime}-p)\xi\right\}
\]%
\begin{align}
&  +4\left(  \delta(\theta-\theta^{\prime}-\pi)+\delta(\theta-\theta^{\prime
}+\pi)\right)  \int\limits_{-\infty}^{\infty}d\xi\chi^{2}(\theta,\xi)\chi
^{2}(\theta^{\prime},\xi+\pi)\nonumber\\
&  \cos\left\{  \frac{2}{\hbar}\left(  p^{\prime}-p+\pi\right)  \xi\right\}
\tag{3.7}\label{3.7}%
\end{align}
Hence, as $\chi(\theta,\xi)$ has a compact support with respect to $\xi$ the
formula (3.7) never gives $2\pi\hbar\delta(\theta-\theta^{\prime}%
)\delta(p-p^{\prime}).$ Consequently, \textit{the axiom} (2.42) \textit{cannot
be satisfied for an arbitrary function on the cylinder}. (Note that this is
always the case if the configuration space $M$ is such that for some point $q$
of $M$ the normal coordinates at $q$ cannot be extended to all the tangent
space $T_{q}(M)$). We must mention here that in the important works [27] the
SW quantizer on the cylinder satisfying the axioms (2.40), (2.41) and (2.42)
has been found. However, this SW quantizer has a disadvantage (which also has
our SW quantizer (3.3)), that is, it does not fulfill the condition:
$\widehat{f}_{W}=f(\widehat{p})$ for arbitrary function $f=f(p)$, which could
be expected for a particle on the circle. The same occurs in the interesting
approach of Alcalde [28] where the notion of SW quantizer is not used. In
fact, as it is known from Ref. [16] the violation of the above condition will
always appear unless we consider a ''quantization'' of the classical
cylindrical phase space.

This quantization in the WUE formalism can be obtained by some limiting
process. Namely, let $\left\{  \chi_{j}\left(  \theta,\xi\right)  \right\}
_{j\in\mathbb{N}}$ be a series of cutoff functions such that for every
$j\in\mathbb{N}$ and every $\theta\in\lbrack-\pi,\pi\lbrack$%
\begin{equation}
0\leq\chi_{j}\left(  \theta,\xi\right)  \leq1,\quad\chi_{j}\left(  \theta
,\xi\right)  =0\text{ for }\xi\notin\text{ }]-\frac{\pi}{2},\frac{\pi}%
{2}[\tag{3.8}\label{3.8}%
\end{equation}
and
\begin{equation}
\lim_{j\rightarrow\infty}\int\limits_{-\frac{\pi}{2}}^{\frac{\pi}{2}}%
d\xi\left(  \chi_{j}\left(  \theta,\xi\right)  \right)  ^{m}f(\xi
)=\int\limits_{-\frac{\pi}{2}}^{\frac{\pi}{2}}d\xi f(\xi)\tag{3.9}\label{3.9}%
\end{equation}
for every $m\in\mathbb{N}$ and every continuous function $f=f(\xi)$ (see
Vladimirov \cite{Vlad}, Section 2.2).

Assuming that the momentum $p=n\hbar,$ $n\in\mathbb{Z},$ using (3.8) and
(\ref{3.9}) one quickly finds that (3.3) leads to
\[
\lim_{j\rightarrow\infty}\langle\varphi_{k}\mid\widehat{\Omega}_{j}%
(k\hbar,\theta)\mid\varphi_{k^{\prime}}\rangle=\frac{1}{\pi}\exp\left\{
i\left(  k^{\prime}-k\right)  \theta\right\}
\]%
\begin{equation}
\lim_{j\rightarrow\infty}\int\limits_{-\frac{\pi}{2}}^{\frac{\pi}{2}}d\xi
\chi_{j}^{2}\left(  \theta,\xi\right)  \exp\left\{  i\left(  k+k^{\prime
}-2n\right)  \xi\right\}  =\langle\varphi_{k}\mid\widehat{\Omega}%
(n,\theta)\mid\varphi_{k^{\prime}}\rangle\tag{3.10}\label{3.10}%
\end{equation}
where $\widehat{\Omega}(n,\theta)$ is the \textit{discrete Stratonovich-Weyl
quantizer for the cylinder} found by Mukunda \cite{Mukunda} and also given in [16,18,19].

Then from (3.7) with (\ref{3.8}) and (\ref{3.9}) we have
\begin{equation}
\lim_{j\rightarrow\infty}Tr\left\{  \widehat{\Omega}_{j}(n,\theta
)\widehat{\Omega}_{j}(n^{\prime},\theta^{\prime})\right\}  =2\pi
\delta_{n,n^{\prime}}\delta(\theta-\theta^{\prime})\tag{3.11}\label{3.11}%
\end{equation}
(Compare with \cite{Pleb2}).

Finally, note that the discrete SW quantizer (3.10) gives: $\widehat{f}%
_{W}=f(\widehat{p})$ for every function $f=f(p)$ as it is expected.

\bigskip

\bigskip

\textbf{Acknowledgments}

This paper was partially supported by CONACYT and CINVESTAV (M\'{e}xico) and
by KBN (Poland). One of us (M.P.) is grateful to the staff of Departamento de
F\'{i}sica at CINVESTAV, (M\'{e}xico, D.F.) for warm hospitality.
\end{document}